\documentclass[%
reprint,
superscriptaddress,
prl,
 amsmath,
 amssymb,
 aps,
 longbibliography
]{revtex4-2}

\usepackage{cancel}
\usepackage{picture}
\usepackage{graphicx}
\usepackage{dcolumn}
\usepackage{bm}
\usepackage[dvipsnames]{xcolor}
\usepackage{soul}
\usepackage{subcaption}
\usepackage{enumitem}
\usepackage{xspace}
\usepackage{booktabs}
\usepackage{lipsum}
\usepackage{siunitx} 
\usepackage[ISO]{diffcoeff}
\usepackage{ragged2e}
\usepackage{mathrsfs}
\usepackage{hyperref}
\usepackage[capitalise]{cleveref}
\usepackage[normalem]{ulem}
\usepackage{amsmath}
\usepackage{wrapfig}
\usepackage{esvect} 
\usepackage{titlesec}
\usepackage{titletoc}
\usepackage{chngcntr}
\usepackage{fontawesome5}
\usepackage{mleftright}
\usepackage[nodayofweek]{datetime}
\usepackage{comment}

\DeclareSIUnit\year{yr}

\DeclareCaptionJustification{justified}{\justifying}
\definecolor{firebrick}{HTML}{B22222}

\hypersetup{
 colorlinks=true, 
 linkcolor=firebrick, 
 citecolor=firebrick, 
 filecolor=firebrick, 
 urlcolor=firebrick 
}

\definecolor{orcid-green}{RGB} {166, 206, 57}
\newcommand{\MYhref}[3][blue]{\href{#2}{\color{#1}{#3}}}%

\titleclass{\mysection}{straight}[\section]
\titleformat{\mysection}[runin]
 {\itshape}{\thesection}{}{}[.---]
\titlespacing{\mysection}{1em}{1em}{0em}

\newcommand\funop[1]{\mathop{{}#1}}
\newcommand{\dd}{\mathop{}\!\mathrm{d}}
\newcommand{\mrm}[1]{\mathrm{#1}}
\newcommand{\usim}{\mathord{\sim}}
\newcommand{\bv}[1]{\mathbf{#1}}
\newcommand{\dv}[2]{\frac{\dd #1}{\dd #2}}
\newcommand{\dvil}[2]{\dd #1/\dd #2}
\newcommand{\uni}[1]{\, \mathrm{#1}}
\newcommand{\mcal}[1]{\mathcal #1}

\DeclareSIUnit\clight{c}

\begin{document}

\title{\boldmath First Search for Ultraheavy Dark Matter Using a Magnetically Levitated Particle}

\author{Dennis G.~Uitenbroek\,\MYhref[orcid-green]{https://orcid.org/0009-0008-3425-6406}{\faOrcid}}
\altaffiliation{Equal contributions}
\affiliation{Leiden Institute of Physics, Leiden University, P.O. Box 9504, 2300 RA Leiden, The Netherlands.}

\author{Dorian W.~P.~Amaral\,\MYhref[orcid-green]{https://orcid.org/0000-0002-1414-932X}{\faOrcid}}
\email{damaral@ifae.es}
\altaffiliation{Equal contributions}
\affiliation{Institut de F\'{\i}sica d'Altes Energies (IFAE), The Barcelona Institute of Science and Technology,
Campus UAB, 08193 Bellaterra (Barcelona), Spain}
\affiliation{Department of Physics and Astronomy, Rice University, MS-315,
Houston, TX, 77005, U.S.A.}

\author{Juehang Qin\,\MYhref[orcid-green]{https://orcid.org/0000-0001-8228-8949}{\faOrcid}}
\altaffiliation{Equal contributions}
\affiliation{Department of Physics and Astronomy, Rice University, MS-315,
Houston, TX, 77005, U.S.A.}
\affiliation{Natural Sciences and Science Education, National Institute of Education,\\ Nanyang Technological University, Singapore 637616, Singapore}

\author{Jurriaan Langendorff}
\affiliation{Leiden Institute of Physics, Leiden University, P.O. Box 9504, 2300 RA Leiden, The Netherlands.}

\author{Andrew Gingerich\,\MYhref[orcid-green]{https://orcid.org/0009-0001-7363-1722}{\faOrcid}}
\affiliation{Department of Physics and Astronomy, Rice University, MS-315,
Houston, TX, 77005, U.S.A.}

\author{Tjerk H.~Oosterkamp\,\MYhref[orcid-green]{https://orcid.org/0000-0001-6855-5190}{\faOrcid}}
\affiliation{Leiden Institute of Physics, Leiden University, P.O. Box 9504, 2300 RA Leiden, The Netherlands.}

\author{Christopher D.~Tunnell\,\MYhref[orcid-green]{https://orcid.org/0000-0001-8158-7795}{\faOrcid}}
\affiliation{Department of Physics and Astronomy, Rice University, MS-315,
Houston, TX, 77005, U.S.A.}

\collaboration{POLONAISE Collaboration}
\noaffiliation

\begin{abstract}
\noindent
We present the first search for ultraheavy dark matter using a magnetically levitated particle. The POLONAISE experiment uses a milligram-scale ferromagnet levitated in a superconducting trap, admitting a force sensitivity of $0.07\,\mrm{fN\,Hz^{-1/2}}$ and resolving impulses as small as $1\,\mrm{TeV}/c$. Treating every candidate impulse as a possible dark matter event, we set optimum-interval upper limits on the neutron coupling $\alpha_n$ for dark matter interacting through a new light mediator. For dark matter masses $10^6\,\mrm{GeV}/c^2\text{--}10^{15}\,\mrm{GeV}/c^2$ and mediators lighter than $30\,\mrm{meV}/c^2$, we exclude couplings as low as $\alpha_n = 3.2\times 10^{-9}$ at $95\%$ confidence level  and set leading constraints on the dark matter–neutron cross section for composite dark matter. Our results extend levitated sensing beyond the mass reach of optical levitation by seven orders of magnitude into the ultraheavy dark matter frontier. \href{https://github.com/PolonaiseExperiment/luhdm.git}
{\faGithub}
\end{abstract}
\maketitle

\mysection{Introduction}

Despite overwhelming astrophysical and cosmological evidence, the nature of dark matter (DM) remains unknown~\cite{Bertone:2004pz,Bertone:2016nfn,Freese:2017idy,Cirelli:2024ssz}. The weakly interacting massive particle (WIMP) has long been the leading candidate, motivating decades of direct detection experiments targeting the $\mrm{GeV}\text{--}\mrm{TeV}$ mass window~\cite{Schumann:2019eaa,LZ:2024zvo,XENON:2025vwd,PandaX:2024qfu}. However, the absence of any confirmed signal in this range has inspired searches beyond the WIMP paradigm. \textit{Ultraheavy} dark matter (UHDM), dark matter with masses exceeding roughly $10^{4}\,\mrm{GeV}/c^2$, has attracted growing interest as an alternative candidate~\cite{Carney:2022gse}, including asymmetric dark matter~\cite{Zurek:2013wia,Wise:2014jva,Wise:2014ola,Krnjaic:2014xza,Gresham:2018anj}, quark nugget dark matter~\cite{Bai:2018vik,Bai:2018dxf,Zhitnitsky:2021iwg}, $Q$-balls~\cite{Kusenko:2001vu}, WIMPzillas~\cite{Chung:1998zb,Kuzmin:1998uv,Kolb:2023ydq}, dark blobs~\cite{Grabowska:2018lnd}, and dark clumps~\cite{Amaral:2025lax}.

Ultraheavy dark matter can interact with Standard Model matter via a long-range, Yukawa-like force mediated by a new light boson. For sufficiently light mediators, this interaction remains coherent across a macroscopic target, producing a force proportional to the number of coupled constituents. Moreover, if the DM itself admits a finite spatial extent, coherence can be maintained over the size of a macroscopic sensor while being lost for nuclear momentum transfers, suppressing the corresponding signal in conventional direct detection experiments~\cite{Coskuner:2018are}. This coherent enhancement has motivated a growing set of searches for long-range DM interactions with macroscopic force sensors~\cite{Hall:2016usm,Monteiro:2020wcb,Tseng:2025rlo,Qin:2025jun,Amaral:2025lax}.

Magnetic levitation has emerged as a promising force sensing platform, featuring ultra-low mechanical dissipation and cryogenic operation that enables force sensitivities of the order of  $10^{-16}\,\mrm{N\,Hz^{-1/2}}$~\cite{Fuchs:2024gravity,Uitenbroek:2026ybe}. The POLONAISE experiment has leveraged this technology to search for signatures of beyond Standard Model physics, including its flagship search for ultralight dark matter~\cite{Amaral:2024uldm}. Magnetic levitation has also been proposed for searches of other ultralight candidates~\cite{Kalia:2024eml,Higgins:2023gwq, Danieli:2026qtx,Peng:2026ffa,Li:2026rty}, novel spin-independent and spin-dependent interactions~\cite{Amaral:2025zgk,Amaral:2026xqz}, high-frequency gravitational waves~\cite{Carney:2024zzk}, and ultraheavy dark matter~\cite{Qin:2025jun}. 

Magnetic levitation is particularly well suited to searches for heavy dark matter, with the coherent enhancement over the sensor significantly amplifying the impulse signal from a passing DM particle. Optical levitation experiments have also recently searched for this signature; however, their up-to nanogram-scale targets limit their sensitivity at DM masses above $\usim10^{8}\,\mrm{GeV}/c^2$~\cite{Monteiro:2020wcb,Tseng:2025rlo}. Probing heavier DM requires larger targets read out with comparable impulse sensitivity and monitored long enough to catch a rare transit---a combination no platform has achieved until now. POLONAISE brings these three elements together: a large milligram-scale mass, sub-femtonewton force sensitivity, and month-scale continuous operation.

In this \textit{Letter}, we present the first experimental search for ultraheavy dark matter using a magnetically levitated particle. We monitor a milligram-scale ferromagnet levitated in a superconducting trap for the momentum kicks delivered by passing dark matter particles. Using the optimum-interval method, we set limits on neutron--philic dark matter interacting via a long-range, Yukawa-type force over the dark matter mass range $4\times 10^{5}\,\mrm{GeV}/c^2\text{--}8 \times 10^{14}\,\mrm{GeV}/c^2$. Our results are sensitive to mediator masses up to $\usim $$30\,\mrm{meV}/c^2$ and apply to any dark matter candidate coupled to neutrons via a long-range force. We recast our results for a concrete composite dark matter model, setting leading constraints on the DM--neutron cross section.
We establish magnetically levitated sensors and the POLONAISE experiment as sensitive new probes of ultraheavy dark matter.

\mysection{Dark Matter Signal}
\label{sec:theory}

Several theories of UHDM posit that dark matter interacts with matter in the visible sector via a long-range force mediated by a new light boson $\varphi$~\cite{Wise:2014jva,Gresham:2018anj,Grabowska:2018lnd}. The potential that describes this force is 
\begin{equation}
 U(r) = \frac{\hbar c\alpha}{r}e^{-r/\lambda}~~\text{with}~~\alpha \equiv \frac{\mcal{N}_n g_n\mcal{N}_d g_d}{4\pi} \equiv \mcal{N}_n \alpha_n\,,
 \label{eq:yukawa}
\end{equation}
where $\alpha$ is the total interaction strength, $\lambda = \hbar / (m_\varphi c)$ is the force range with $m_\varphi$ the mediator mass, and $r$ is the relative distance between the interacting particles. As in the optical levitation experiments carried out by Refs.~\cite{Monteiro:2020wcb,Tseng:2025rlo}, we assume the DM to be neutron--philic and to be plausibly composed of more fundamental dark constituents. The interaction strength can then be written in terms of the couplings to neutrons and dark constituents, $g_n$ and $g_d$, having respective charge numbers $\mcal{N}_n$ and $\mcal{N}_d$. Ultimately, we report our results with respect to the total DM--neutron coupling $\alpha_n \equiv \mcal{N}_d g_d g_n / (4\pi)$, expressing the strength of the force imparted by the dark matter particle per neutron.

\begin{figure}[t!]
 \centering
 \includegraphics{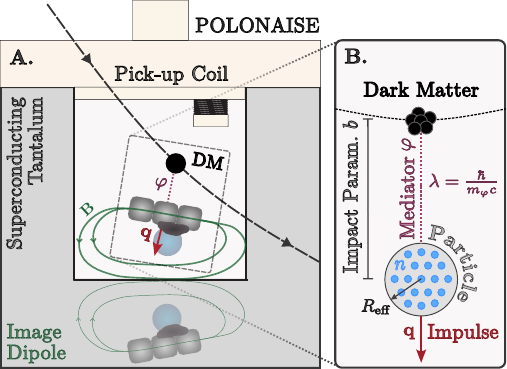}
 \caption{\textbf{A.}~The POLONAISE experiment. A $0.356 \uni{mg}$ magnetic particle composed of three $\mrm{Nd_2Fe_{14}B}$ cubes and a glass sphere is levitated inside a Type-I superconducting tantalum trap. The motion of the magnet is read out by a superconducting pick-up coil coupled to a DC SQUID. A passing dark matter particle imparts an impulse $q$ on the magnet. Further details of the apparatus, including schematics of the vibration isolation system and cryostat, are given in Refs.~\cite{Fuchs:2024gravity,Uitenbroek:2026ybe}. \textbf{B.}~Inset showing the dark matter particle interacting with the magnetic particle via the neutron--philic force carrier $\varphi$ of mass $m_\varphi$, mediating a Yukawa-like force. The force has range $\lambda$, while the dark matter passes by with impact parameter $b$. The magnetic particle is modeled as a uniform sphere of effective radius $R_\mrm{eff} \approx 260 \uni{\mu m}$.}
 \label{fig:exp-setup}
\end{figure}

As a UHDM particle passes by the detector, it exerts a force on the levitated particle, imparting an impulse $q$ as shown in \cref{fig:exp-setup}. The expected number of these impulses follows from the differential rate of DM interactions,
\begin{equation}
 \left\langle \dv{R}{q} \right\rangle = n_\mrm{DM} \int f_\mrm{det}(\bv{v}) v \dv{\sigma}{q}(q, v) \dd^3\bv{v}\,,
 \label{eq:diff_rate}
\end{equation}
where $n_\mrm{DM} \equiv f_\mrm{DM}\rho_\mrm{DM} / m_\mrm{DM}$ is the number density of DM particles in our local neighborhood, with $\rho_\mrm{DM} = 0.3 \uni{GeV\,cm^{-3}}$~\cite{deSalas:2020hbh,ParticleDataGroup:2024cfk} and $f_\mrm{DM} \leq 1$ the fraction of DM that is composed of our candidate. The two main ingredients entering \cref{eq:diff_rate} are the probability density function $f_\mrm{det}(\bv{v})$ of the dark matter velocities at the detector and the differential scattering cross section $\dvil{\sigma}{q}$.

For DM particles to reach our detector, they must first traverse through the atmosphere, where they can scatter with air molecules and lose momentum. The distribution of DM velocities at the detector is governed by $f_\mrm{det}(\bv{v}|\alpha_n, \lambda)$, which is dependent on both the strength and the range of the force. To compute it, we begin by sampling velocities from the standard halo model in the lab frame, which captures the local unattenuated distribution of DM velocities~\cite{Lewin:1995rx}. We then evolve these velocities through the momentum-transfer equation governed by the momentum-transfer cross section. For weak coupling strengths or short force ranges, the velocity distribution at the detector is largely unmodified from the halo model as scatters are weak and unlikely to occur. However, for stronger couplings $\alpha_n \gtrsim 10^{-3}$, the distribution can be highly distorted towards lower velocities, potentially decelerating DM particles to speeds below our momentum threshold~\cite{supp_mat}.

Once inside the detector, a DM particle can scatter with our sensor, transferring a momentum $q$ to a single sensor read-out axis that is controlled by the differential scattering cross section $\dvil{\sigma}{q}$. We evaluate this quantity classically in the small-angle scattering regime, which is valid for the weak interaction strengths we consider. Modeling the levitated particle as a sphere of effective radius $R_\mrm{eff}$ with uniform neutron density, we compute the transferred momentum  analytically for DM trajectories with impact parameters $b \geq R_\mrm{eff}$. We conservatively consider only external trajectories since the highest momentum transfers occur at grazing incidences to the detector $b \sim R_{\mrm{eff}}$; internal trajectories contribute only an order unity correction to the cross section. We compute the differential cross section for impact parameters from $R_\mrm{eff}$ to the largest value producing this momentum transfer~\cite{supp_mat}. Throughout our analysis, we treat the DM as point-like, which is valid provided that its spatial extent is small compared to both the force range and the sensor.

\mysection{Experiment}

Originally developed to measure small-scale gravity and recently used to perform the first search for ultralight dark matter using a magnetically levitated sensor, the POLONAISE apparatus provides the near-continuous monitoring required of a UHDM detector~\cite{Fuchs:2024gravity,Amaral:2024uldm}. We show a schematic of our setup and an illustration of the type of DM event we search for in \cref{fig:exp-setup}. In its present configuration, we levitate a sub-millimeter $\mrm{Nd_2Fe_{14}B}$ permanent magnet in a Type-I superconducting tantalum trap. A $0.20\,\mrm{mm}$ glass sphere is glued to the magnet to break rotational symmetry, giving the magnetic particle a total levitated mass of $m_\mrm{mag} = 0.356\uni{mg}$ and a neutron charge number of $\mcal{N}_n \simeq m_\mrm{mag}/(2 u) \approx 1.07\times10^{20}$, where $u \approx 1.67 \times 10^{-27}\,\mrm{kg}$ is the atomic mass unit. The effective radius of the particle is $R_\mrm{eff} \approx 260\uni{\mu m}$, obtained geometrically by summing the volumes of the subcomponents of the magnet into one equivalent sphere.

The levitated magnet exhibits six eigenmodes (three translational and three librational), all of which are continuously monitored~\cite{Uitenbroek:2026ybe}.
We take the modes at the frequencies $f_0$ of $51.2\uni{Hz}$, $59.5 \uni{Hz}$, and $94.9\uni{Hz}$ to be the translational eigenmodes, indexing them by ascending frequency. The search presented here uses mode 1, which has quality factor $Q=2.25\times 10^6$; the remaining two modes are analyzed as cross-checks and are included in the data release. The noise floor of this mode near $f_0$ is $S_{FF}^{1/2} = 0.07\,\mrm{fN\,Hz^{-1/2}}$.

We detect the motion of the magnet using a superconducting pick-up coil. As the magnet moves, the magnetic flux threading the coil changes, inducing a current in a circuit comprising the pick-up coil, a calibration loop, and the input coil of a two-stage DC SQUID~\cite{Uitenbroek:2026ybe,Fuchs:2024gravity}.
We use the calibration loop to fix the energy coupling between the detection circuit and each mechanical mode. Environmental vibrations transmitted to the trap dominate our noise budget. We reduce them by suspending the trap from a multi-stage cryogenic mass-spring system inside a dry dilution refrigerator, attenuating environmental vibrations by $110\text{--}130\uni{dB}$ at the mode frequencies~\cite{Uitenbroek:2026ybe}. Vibrations are further attenuated by resting the cryostat on a $25$-metric-ton concrete block that is supported by pneumatic dampers and mounting the pulse-tube cooler and vacuum pumps on a separate frame, such that they are coupled to the cryostat only through edge-welded bellows and soft copper braid~\cite{Fuchs:2024gravity}.

We recorded data between 22 December 2025 and 21 January 2026. Based on our logbooks, we analyze only night-time data, $19{:}00$ to $07{:}00$ local time, when environmental noise is expected to be lowest. We exclude 19 and 20 January, when work on the still-suspension and a deliberate vibration test disturbed the apparatus, as well as the periods when the calibration drive was on. This selection leaves a livetime of $T_\mrm{obs} = 219.66\uni{h}$ ($9.15\uni{days}$), $31\%$ of the $30$-day run.

\mysection{Analysis}

We search for impulses in the demodulated sensor output. The readout is
demodulated at the three mode frequencies simultaneously with a digital lock-in and low-pass filter, and the ends of each record are trimmed to avoid edge effects. The three modes are analyzed independently and no coincidence between them is required. We analyzed mode~1 at the outset of the analysis, and nothing that emerged later favored a change strongly enough to justify the trial-factor penalty of re-selecting a mode. Modes~1 and~2 have comparable noise floors. Mode~1 yields the fewest transients, and its background is the most clearly diurnal, rendering it the cleanest candidate sample following our night-time selection; mode~2, carried through the same chain as a cross-check, gives comparable  results. All three modes are included in the data release~\cite{luhdm_datarelease}.

\begin{figure}[t]
 \centering
 \includegraphics{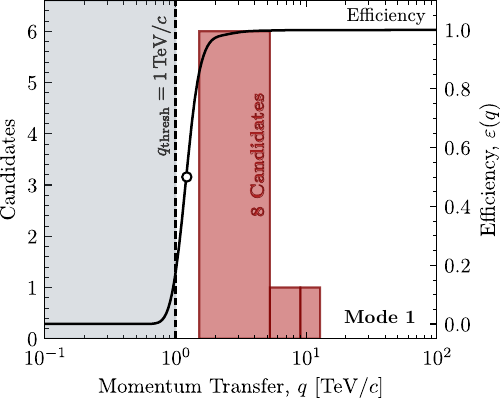}
 \caption{The reconstructed impulse spectrum of the 8 candidates surviving the impulse selection in mode 1 (histogram, left axis), with the segment-averaged detection efficiency $\varepsilon(q)$ of this selection overlaid (right axis)~\cite{supp_mat}. The dashed line marks the selection threshold $q_\mrm{thresh} = 1\uni{TeV}/c$, where the efficiency is $17\%$. The white circle marks the $50\%$ efficiency point, where $q \approx 1.21\uni{TeV}/c$.}
 \label{fig:data-spectrum}
\end{figure}

\begin{figure*}
    \centering
\includegraphics{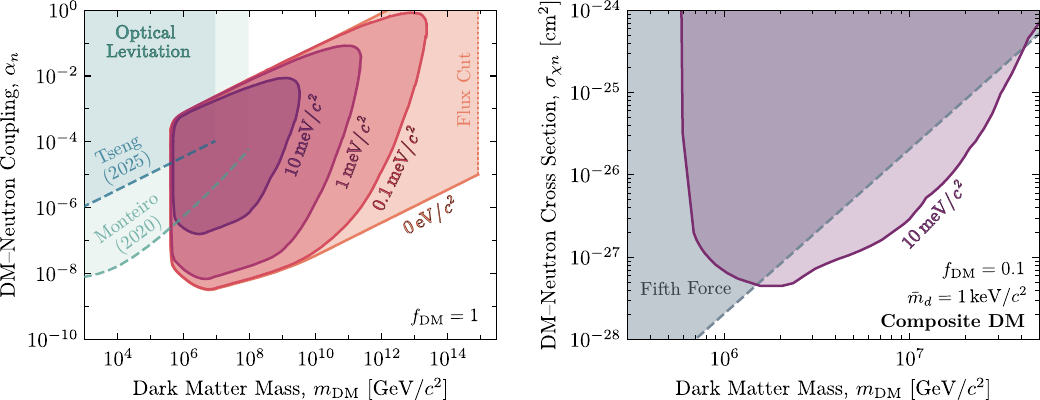}
    \caption{\textbf{Left:} The $95\%$ CL limits on the DM--neutron coupling $\alpha_n$ with dark matter mass $m_\mrm{DM}$ for the mediator masses $m_\varphi$ shown and a dark matter fraction $f_\mrm{DM} = 1$. The limits are capped at the DM mass that produces a negligible flux within the superconducting trap, shown by the dotted line. The dashed curves show the existing limits from dark matter searches employing optical levitation strategies for a massless mediator~\cite{Monteiro:2020wcb,Tseng:2025rlo}.
     \textbf{Right:} The limit for the mediator mass $m_\varphi = 10\,\mrm{meV}/c^2$  recast to the composite benchmark model on the DM--neutron cross section $\sigma_{\chi n}$ for a dark matter fraction $f_\mrm{DM} = 0.1$. The composite DM particle is assumed to have constituents of mass $\bar{m}_d = 1\,\mrm{keV}/c^2$. The dashed line shows the leading limit from a fifth force torsion balance experiment~\cite{Lee:2020zjt,supp_mat}.}
    \label{fig:limits}
\end{figure*}

The demodulated stream is split into
one complex sample every $2.0\uni{s}$, chosen such that the differences between subsequent samples are approximately independent. Each initially $\usim10 \uni{min}$ run then yields $272$ samples and $271$ first differences, spanning $542\uni{s}$ of effective livetime after trimming and decimation. An instantaneous impulse then appears as a persistent step in the complex amplitude
since the ringdown times of the modes, given by $1/\gamma \approx 3000\text{--}7000\uni{s}$ with $\gamma \equiv \omega_0/Q$ the damping factor and $\omega_0 \equiv 2 \pi f_0$, far exceed the sample spacing. A step is equivalent to an excess in one, or at most two adjacent, first differences. 

To quantify whether an impulse is present, we define the test statistic $\mcal{T}$ based on the profile-likelihood ratio. We build the likelihood from the complex first differences under a kick shared between the two points, testing each adjacent pair against the impulse-free hypothesis. We include nuisance parameters for the fraction of the kick shared between the two adjacent differences and for the noise power of the segment. The null distribution of $\mcal{T}$ is known in closed form, and we confirm it via numerical simulation of noise-only data~\cite{supp_mat}. We select excursions with $\mcal{T} > 100$ as impulse candidates, chosen so that Gaussian-noise accidentals are negligible: the exact null predicts $4.4\times10^{-15}$ across the whole search, and the search is limited instead by non-Gaussian instrumental transients. Further details of our statistical analysis are given in Ref.~\cite{supp_mat}.

Each contiguous excursion above threshold defines one candidate impulse. The amplitude of this impulse is converted to a momentum via the mode's displacement calibration, relating the demodulated voltage picked up by the calibration loop to the magnet's motion, velocity, and finally momentum~\cite{supp_mat}. This chain carries an $8.5\%$ scale uncertainty from the displacement calibration, which we do not
propagate, reporting our limits at the nominal scale instead~\cite{Uitenbroek:2026ybe}. Mode 1 yields eight candidate events with impulses
spanning $1.52\uni{TeV}/c\text{--}12.8\uni{TeV}/c$ with median of $3.1\uni{TeV}/c$, having signal-to-noise ratios spanning $12.0$--$34.3$, well above the selection threshold.
The transient selection reaches $50\%$ efficiency at $1.21\uni{TeV}/c$; however, we set the momentum threshold of the search at $q_\mrm{thresh} = 1\uni{TeV}/c$, where the efficiency is $17\%$. This is because we expect systematic uncertainties in the efficiency to be important below this point~\cite{supp_mat}. Compared to the standard quantum limited momentum transfer $\Delta p_\mrm{SQL} \sim (\hbar m_\mrm{mag} \omega_0
)^{1/2} \approx 0.2\,\mrm{GeV}/c$, our experiment remains in the classical readout regime and is limited by classical noise sources. The rate in \cref{eq:diff_rate} is integrated above $q_\mrm{thresh}$. The candidate spectrum and efficiency curve for mode 1 are shown in \cref{fig:data-spectrum}.

The eight candidates that survive our selection are consistent with instrumental
transients: the population is non-stationary and correlates with laboratory activity. However, with no reliable model to capture the spectrum of these transients, we cannot subtract them as background. Therefore, we cannot claim a discovery and must instead treat every candidate as a potential dark matter event, deriving upper limits on the signal strength. To compute our limits, we first compute the number of expected counts by folding our computed efficiency curve and livetime into the differential rate given in \cref{eq:diff_rate}. For each hypothesis on $(\alpha_n, \lambda, m_\mrm{DM})$, we then derive our limits using the optimum-interval
method~\cite{Yellin:2002xd,optimum_interval}, which bounds the signal from the most
anomalously empty interval of the observed spectrum. The construction over-covers, so
our exclusions are conservative at the $95\%$ confidence level (CL)~\cite{supp_mat}.

\mysection{Results and Discussion}

We show the $95\%$ CL limits on the DM--neutron coupling $\alpha_n$ with dark matter mass in \cref{fig:limits} (left) assuming that our DM candidate composes the entirety of DM. We consider a massless mediator ($\lambda \rightarrow \infty$) and the mediator masses $0.1\,\mrm{meV}/ c^2$ ($\lambda \approx 2\,\mrm{mm}$), $1\,\mrm{meV}/c^2$ ($\lambda \approx 200\,\mrm{\mu m}$), and $10\,\mrm{meV}/c^2$ ($\lambda \approx 20\,\mrm{\mu m}$). Our best limits are derived for mediator masses $m_\varphi \ll 1\,\mrm{meV}/c^2$, corresponding to force ranges larger than the effective radius of our magnetic particle $\lambda \gg R_{\mrm{eff}}$, leading to the total coherent enhancement over our sensor. For heavier mediators, the force between the DM particle and the magnet is suppressed due to both the loss of coherence and the exponential Yukawa factor for impact parameters $b > \lambda$. This weakens our limits, and we lose sensitivity for mediators heavier than $m_\varphi \gtrsim 30\,\mrm{meV}/c^2$ ($\lambda \lesssim 7\,\mrm{\mu m}$)~\cite{supp_mat}. 

For a massless mediator, we best constrain $\alpha_n \gtrsim 3.2\times10^{-9}$ at the DM mass $m_\mrm{DM} \approx 5 \times10^{6}\,\mrm{GeV}/c^2$, with the limit rising as $\alpha_n \propto m_\mrm{DM}^{1/2
}$ due to the decreasing DM flux. 
Our limit extends over the DM mass range $4\times10^{5}\,\mrm{GeV}/c^2\text{--}8\times10^{14}\,\mrm{GeV}/c^2$, widening the DM mass window probed by levitation experiments by approximately seven orders of magnitude~\cite{Monteiro:2020wcb,Tseng:2025rlo}. This reach is a consequence of the milligram-scale masses that magnetic levitation affords us over the nanogram masses accessible to optical traps, as the coupling reach scales as $\alpha_n \propto q_\mrm{thresh} / \mcal{N}_n$ for a fixed rate. While this enhancement is partially offset by our correspondingly higher momentum threshold, our net reach is improved by $\usim 10^2$ compared to Ref.~\cite{Monteiro:2020wcb}, reporting a threshold of $\usim 0.1\,\mrm{GeV}/c$.

Our limits form closed contours in the coupling and the DM mass. For all mediator masses, our excluded regions begin at $m_\mrm{DM} 
\sim q_\mrm{thresh}/(v_\mrm{esc}+v_\odot) \approx 4 \times 10^{5} \uni{GeV}/c^2$. This is the lowest DM mass that can impart our threshold momentum of $1\,\unit{TeV}/c$ when traveling
at the maximum speed with which halo particles reach the laboratory, with $v_\mrm{esc} \approx 544\,\mrm{km\,s^{-1}}$ the galactic escape speed~\cite{Smith:2006ym,Lewin:1995rx} and $v_\odot \approx 245\,\mrm{km\,s^{-1}}$ the solar speed in the galactic rest frame~\cite{schonrich2010local}. In contrast, optical levitation searches are sensitive to masses as low as $m_\mrm{DM} \sim 1\,\mrm{GeV}/c^2$ owing to their lower thresholds~\cite{Tseng:2025rlo}. 

From above, the contours are bounded by atmospheric attenuation effects, whereby a large enough coupling stops DM particles before they can reach our detector. For massive mediators, these contours then close in DM mass where the effect from this falling number density meets that of the saturating cross section. For the massless mediator, we truncate our limit at the DM mass above which fewer than three particles within
the $\usim 10\,\mrm{cm}$-scale trap are expected during our exposure. We claim no exclusion above this mass: there, the expected number of close transits during the exposure falls below a few, so any limit would rest on a handful of encounters rather than on a rate~\cite{supp_mat}.

Our results apply to any neutron--philic DM particle interacting via a generic Yukawa potential and can therefore be recast in terms of a concrete DM model. Following Refs.~\cite{Monteiro:2020wcb,Tseng:2025rlo}, we consider a composite DM particle of total mass $m_\mrm{DM}$, existing as a spatially extended bound state of more fundamental constituents of mass $\bar{m}_d < m_\mrm{DM}$ and interacting with neutrons via the mediator $\varphi$. Such a composite model is well motivated by theories positing that DM exists as asymmetric dark matter~\cite{Zurek:2013wia,Wise:2014jva,Wise:2014ola,Krnjaic:2014xza,Gresham:2018anj}, quark nugget dark matter~\cite{Bai:2018vik,Bai:2018dxf,Zhitnitsky:2021iwg}, and dark blobs~\cite{Grabowska:2018lnd}.

For this model, we quote an effective DM–neutron cross section of $\sigma_{\chi n} \equiv 4 \pi (\hbar c)^2 \mu_{\chi n}^2 \alpha_n^2 / (m_\varphi^2 +q_0^2)^2$, where $\mu_{\chi n} \simeq m_n$ is the reduced DM--neutron mass and $q_0 \equiv m_n v_0$ is a reference momentum, with $v_0 = 10^{-3}c$~\cite{Digman:2019wdm}. We assume the heaviest mediator for which we draw a contour, $m_\varphi = 10\uni{meV}/c^2$; its force range is the shortest we consider and the fifth force bounds are correspondingly the weakest~\cite{Adelberger:2003zx}. As in Refs.~\cite{Monteiro:2020wcb,Tseng:2025rlo}, we saturate the coupling to $g_d = 1$, near the perturbative limit, and we take the constituent mass to be $\bar{m}_d = 1 \uni{keV}$. This implies that the characteristic size of the DM bound state is smaller than both the force range, $\lambda \approx 20 \uni{\mu m}$, and the effective radius of the magnet, justifying the point-like DM assumption we make throughout our work~\cite{supp_mat}.

We show the $95\%$ CL limits on the DM--neutron cross section $\sigma_{\chi n}$ with dark matter mass in \cref{fig:limits} (right). We assume that our composite DM candidate forms a sub-component of the DM at the level of $f_\mrm{DM} = 0.1$, allowing us to neglect the stringent self-interaction bounds~\cite{Coskuner:2018are,Gresham:2018anj}. The leading constraints on our benchmark DM model arise from a fifth force torsion balance experiment~\cite{Lee:2020zjt,supp_mat}; while DM direct detection experiments place leading constraints on the spin-independent WIMP--nucleon cross section~\cite{XENON:2025vwd,PandaX:2024qfu}, the form factor from the finite size of the DM particle suppresses the nuclear recoil signals due to a significant loss of coherence~\cite{Coskuner:2018are,supp_mat}. We place the leading limit on this DM candidate over the mass range $2 \times 10^{6}\,\mrm{GeV}/c^2\text{--}4\times10^{7}\,\mrm{GeV}/c^2$, constraining cross sections down to $\sigma_{\chi n} = 4.5\times10^{-28}\uni{cm^2}$ and improving existing bounds by an order of magnitude at the DM mass  $9\times 10^{6}\uni{GeV}/c^2$.

\mysection{Conclusions}
Using the POLONAISE experiment, we performed the first search for ultraheavy dark matter with a magnetically levitated particle. We considered a neutron--philic DM candidate interacting via a long-range, Yukawa-like force mediated by a new light boson. 
Monitoring the translational motion of a $0.356\,\mathrm{mg}$ ferromagnet at a force noise of $0.07\,\mrm{fN\,Hz^{-1/2}}$ over $9.15$ days of livetime, we reconstructed eight candidate impulses above a threshold momentum of $1\,\mathrm{TeV}/c$, conservatively treating every candidate as a potential dark matter signal. We set upper limits on the DM coupling strength using the optimum-interval method over the DM masses $4\times10^{5}\,\mrm{GeV}/c^2\text{--}8\times10^{14}\,\mrm{GeV}/c^2$, covering nine orders of magnitude and extending the DM mass window probed by levitation experiments by seven orders of magnitude. We derived constraints for mediators of masses $m_\varphi \lesssim 30\,\mrm{meV}/c^2$, best excluding DM--neutron couplings $\alpha_n \gtrsim 3.2\times10^{-9}$ at the $95\%$ confidence level. Finally, we recast our limits in terms of a concrete, composite DM model, placing a leading constraint on the DM--neutron cross section of $\sigma_{\chi n} \lesssim$ $4.5\times10^{-28}\,\mrm{cm^2}$ for our benchmark candidate.
Milligram-scale magnetic levitation and nanogram-scale optical levitation now probe complementary dark matter mass windows with comparable coupling reach, and the rapid advance of levitated quantum sensing promises to widen both.

\begin{acknowledgments}
We would like to thank Daniel Carney and David Moore for useful discussions regarding the differential cross section, as well as Rafael Lang, Gerard Higgins, and Hendrik Ulbricht for comments on the manuscript. DA has been supported by ERC grant ERC-2024-SYG 101167211 by the European Union, CT by NSF CAREER 2046549, JQ by DOE AI4HEP, and TO/DU/JL by the EU Horizon Europe EIC Pathfinder project QuCoM (10032223).
Views and opinions expressed are however those of the author(s) only and do not necessarily reflect those of the European Union, European Research Council Executive Agency, or other awarding body. Neither the European Union nor the granting authority can be held responsible for them.
\end{acknowledgments}

\section*{Data Availability}
The data and the analysis code that produce every result in this work are
available in Refs.~\cite{luhdm_datarelease,optimum_interval}.

\bibliography{biblio}

\mysection{End Matter: AI-Usage Statement}
We used AI assistance in this work. Anthropic's Claude Opus 4.8 was used to refactor code, including the authors' implementation of the optimum-interval method used on XENON100, and to convert Jupyter notebook prototypes into scripts. Claude Fable 5 was used for copyediting, consistency checking, and numerical cross-checking of the manuscript against the data release. A human author reviewed and takes responsibility for every result and every statement in this work, and no AI system is an author.

\clearpage

\appendix

\setcounter{secnumdepth}{3}

\onecolumngrid

\begin{center}
\large
\textbf{
 \textit{Supplemental Material:} \\ First Search for Ultraheavy Dark Matter Using a Magnetically Levitated Particle
 }

\vspace{1.75ex}

Dorian W.~P.~Amaral, Dennis G.~Uitenbroek, Juehang Qin, Jurriaan Langendorff, Andrew Gingerich, Tjerk H.~Oosterkamp, and Christopher D.~Tunnell

\vspace{2ex}

\end{center}

\twocolumngrid

\section{The Dark Matter Impulse Signal}
\label{sec:signal-model}

We search for impulses delivered to our magnetic particle due to passing DM particles. This is controlled by the differential rate given in \cref{eq:diff_rate}, which tells us the number of DM particles transferring a given momentum $q$ per unit time. The two main ingredients in this calculation are the velocity probability density function at the detector $f_\mrm{det}(\bv{v})$ and the differential scattering cross section $\dvil{\sigma}{q}$.

\subsection{The Detector Velocity Distribution, \boldmath$f_\mrm{det}$}

Before DM particles can scatter with our sensor, they must first make it to the detector. Dark matter particles must therefore  travel through the atmosphere, where they can scatter with air molecules and attenuate their velocities. This means that the distribution of DM velocities at the detector can be distorted relative to that of the DM halo due to atmospheric attenuation effects, modifying the expected DM scattering rate. 

To compute the velocity probability density function at the detector $f_\mrm{det}(\bv{v})$, we begin from the DM halo function. We assume the standard halo model (SHM)~\cite{Lewin:1995rx}: an isotropic and isothermal Maxwell-Boltzmann distribution truncated at the galactic escape velocity $v_\mrm{esc} \approx 544\,\mrm{km\,s^{-1}}$~\cite{Smith:2006ym,Lewin:1995rx}. In the galactic rest frame, DM particles therefore travel at velocities $\bv{u}$ governed by 
\begin{align}
    &f_\mrm{gal}(\bv{u}) \dd^3\bv{u} \equiv \frac{1}{N_0}e^{-|\bv{u}|^2/v_0^2}\funop{\Theta(v_\mrm{esc} - |\bv{u}|)}\dd^3\bv{u}\,,\\
    &\text{with }N_0 \equiv (\sqrt{\pi} v_0)^3\left[\mrm{erf}\left(\frac{v_\mrm{esc}}{v_0}\right) - \frac{2}{\sqrt{\pi}} \frac{v_\mrm{esc}}{v_0}e^{-v_\mrm{esc}^2 / v_0^2}\right]\,, \nonumber
\end{align}
where $v_0 \approx 220\,\mrm{km\,s^{-1}}$ is the local galactic circular rotation speed~\cite{Evans:2018bqy}. 

In the lab frame, which moves at a velocity $\bv{v}_\mrm{lab}$ relative to the galactic rest frame, the relevant boosted DM velocity distribution is $f_\mrm{lab}(\bv{v}) \equiv f_\mrm{gal}(\bv{v} + \bv{v}_\mrm{lab})$. In general, $\bv{v}_\mrm{lab}$ is time dependent, containing contributions from the solar velocity around the galaxy, the Earth velocity around the Sun, and the rotation of the Earth. We ignore this time dependence in this work and set $\bv{v}_\mrm{lab} \simeq \bv{v}_\odot$, the velocity of the Sun relative to the galactic rest frame, with $|\bv{v}_\mrm{\odot}| \approx 245\,\mrm{km\,s^{-1}}$~\cite{Lewin:1995rx,schonrich2010local}. Integrating over angles, we then find
\begin{equation}
\begin{split}
    f_\mrm{lab}(v)\dd v &= \frac{1}{N_0} \frac{\pi v_0^2}{v_\odot} v \left\{\exp\left[-\left(\frac{v - v_\odot}{v_0}\right)^2\right]\right. \\
    &- \left.\exp\left[-\left(\frac{\mrm{min}(v + v_\odot, v_\mrm{esc})}{v_0}\right)^2\right]\right\} \dd v\,.
\end{split}
\label{eq:shm}
\end{equation}
This probability density function encodes the distribution of DM speeds we should expect at the detector; however, it ignores the effect that the atmosphere can have on the momenta of the incoming particles.

Dark matter particles traversing the atmosphere scatter with air molecules, slowing them down as they travel to the detector. This modifies
the distribution of velocities arriving at the detector
from the halo function given in \cref{eq:shm}.  To compute this distribution, we must
know how an initial DM velocity evolves through the atmosphere. 

For a DM particle traveling with velocity $v$ that scatters with some molecular species in the atmosphere, the average change in its momentum due to many scattering events is $\langle \Delta p \rangle = (\sigma_T / \sigma) p$, where $\sigma$ is the total scattering cross section and $p \equiv \mu v$, with $\mu$ the reduced mass of the system. The quantity $\sigma_T$ is the momentum-transfer cross section
\begin{equation}
    \sigma_T \equiv \int (1 - \cos \theta) \dv{\sigma}{\Omega} \dd \Omega\,,
\end{equation}
where $\theta$ is the scattering angle in the center-of-mass frame, describing the average momentum transferred due to a collision. For a Yukawa-like repulsive interaction in the weak-coupling limit, we may write it as 
~\cite{khrapak2004momentum,Colquhoun:2020adl}
\begin{equation}
\begin{split}
 \sigma_T &\simeq 2 \pi \lambda^{2} \beta^{2}\ln\left( 1 + \frac{1}{\beta^{2}} \right) \\
 &\simeq  (\hbar c)^2\frac{4 \pi\alpha_\mrm{mol}^2}{\mu^2 v^4} \ln\left(\frac{1}{\hbar c}\frac{\lambda \mu v^2}{\alpha_\mrm{mol}}\right)\,,
 \end{split}
\end{equation}
where we have defined the coupling coefficient $\beta \equiv \hbar c\alpha_\mrm{mol} / (\lambda \mu v^2)$, with $\beta \ll 1$ for all of our analysis. The DM--molecule coupling is $\alpha_\mrm{mol} \equiv \mcal{N}_n^\mrm{mol} \alpha_n$, where $\mcal{N}_n^\mrm{mol}$ is the number of neutrons in an atmospheric species.

\begin{figure}[t]
 \centering
 \includegraphics{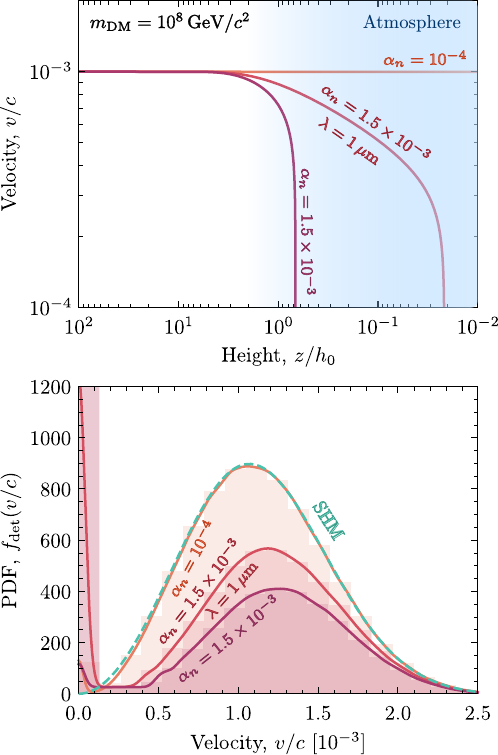}
 \caption{The effect of atmospheric attenuation on a dark matter particle of mass $m_\mathrm{DM} = 10^{8}\,\mathrm{GeV}/c^{2}$. Three benchmark points are considered in the DM--neutron coupling $\alpha_n$ and force range $\lambda$, with $\lambda = 1\,\mathrm{\mu m}$ for one benchmark and $\lambda \rightarrow \infty$ for the remainder. \textbf{Top:} The evolution of a dark matter particle's velocity $v(z)$ as it travels through the atmosphere with an initial velocity $v = 10^{-3}c$. The atmospheric density profile is overlaid. Only the benchmark point with $\alpha_n = 10^{-4}$ allows this dark matter particle to make it to our detector. \textbf{Bottom:} The dark matter speed probability density function at the detector $f_\mrm{det}(v)$. Greater attenuation leads to more particles piling up at low velocities.}
 \label{fig:atmosphere}
\end{figure}

The average momentum loss rate can be written in terms of $\sigma_T$ as~\cite{Colquhoun:2020adl}
\begin{equation}
\frac{\langle\dot{p}\rangle}{\langle p \rangle} = n_\mrm{atm}(z) \sigma v \frac{\langle \Delta p \rangle}{m_\mrm{DM} v} =  \frac{\mu}{m_\mrm{DM}}n_\mrm{atm}(z) \sigma_T v\,,
\label{eq:mom_rate}
\end{equation}
where $n_\mrm{atm}(z)$ is the number density of atmospheric molecules with height $z$ and, for the heavy DM we consider in this work, $\mu \simeq m_\mrm{mol}$, with $m_\mrm{mol}$ the mass of an atmospheric molecule. Assuming that DM particles travel directly towards our detector from the zenith, we can further write $v = -\dot{z}$. From \cref{eq:mom_rate}, the average velocity then evolves according to
\begin{equation}
 \dv{v}{z} = \frac{m_\mrm{mol}}{m_{\mrm{DM}}} n_{\mathrm{atm}}(z) \sigma_{T} v \,.
 \label{eq:v_evolve}
\end{equation}

We solve this equation numerically to propagate an initial DM velocity entering the atmosphere to a final velocity at the surface of the Earth. We take the atmosphere to be predominantly composed of $\mrm{N}_2$ molecules, such that $\mcal{N}_n^\mrm{mol} = 14$ with $m_\mrm{mol} \simeq 28 u \approx 26\,\mrm{GeV}/c^2$. We assume the number density to follow an exponential profile in height as $n_\mrm{atm}(z) = n_0 e^{-z/h_0}$, with $n_0 \approx 2.4 \times 10^{25}\,\mrm{m^{-3}}$ the average number density in the troposphere and $h_0 \approx 8.5\,\mrm{km}$ its effective height~\cite{Tseng:2025rlo}. We show the resulting velocity evolution with height for a DM particle with mass $m_\mrm{DM} = 10^{8}\,\mrm{GeV}/c^2$ and initial velocity
$v_i = 10^{-3}c$ for varying choices of $\alpha_n$ and $\lambda$ in \cref{fig:atmosphere} (top). Larger values for either of these quantities result in greater attenuation, with only one of our shown benchmark trajectories making it through to the detector.

We also use \cref{eq:v_evolve} to obtain the speed probability density function of DM particles arriving at the detector, $f_\mrm{det}(v)$. We begin by sampling speeds from $f_\mrm{lab}(v)$, given in \cref{eq:shm}, representing the initial DM particle speeds entering the upper atmosphere. Using \cref{eq:v_evolve}, we then evolve these speeds assuming zenith-pointing trajectories to obtain the distribution of final velocities. A kernel density estimator is used on the resulting distributions, allowing us to compute them for arbitrary speeds. We show our results in \cref{fig:atmosphere} (bottom), demonstrating that weak couplings do not significantly distort the detector distributions from the prediction of the SHM. Larger couplings lead to greater attenuation, resulting in slower DM trajectories around the detector, and a large peak at zero velocities where DM particles can be completely stopped by the atmosphere.

\subsection{The Differential Cross Section, \boldmath$\dvil{\sigma}{q}$}
\label{subsec:cross_section}

Once a DM particle has made it to the detector, it can scatter with our sensor. The probability that it scatters to produce a given momentum transfer $q$ is encoded in the differential scattering cross section. To compute it, we must derive the force stemming from the Yukawa potential in \cref{eq:yukawa}, the resultant momentum transfer, and finally the differential cross section projected onto one of the readout axes of the sensor. 

Consider a passing, point-like DM particle at position $\bv{r}(t)$ interacting with our spatially extended magnetic particle, which is placed at the origin. The point-like DM treatment is valid as long as the size of the DM particle is smaller than both the force range $\lambda$ and the magnetic particle effective radius $R_\mrm{eff}$. Integrating the Yukawa potential given in \cref{eq:yukawa} over the neutron charge in the magnetic particle gives the total potential energy
\begin{equation}
 U(\bv{r}) = \hbar c \alpha \int n_n(\bv{r'})\frac{ e^{-|\bv{r} - \bv{r'}|/\lambda}}{|\bv{r} - \bv{r'}|}\dd^3\bv{r'}\,,
\end{equation}
where $n_n(\bv{r}')$ is the neutron number density of the magnetic particle and $\alpha = \mcal{N}_n\alpha_n$ is the total DM--magnet coupling. For a spherically symmetric neutron density, 
\begin{equation}
 n_n(r) =
\begin{cases}
3 / (4\pi R_\mrm{eff}^3)\quad&\text{if $r \leq R_\mrm{eff}$}\,,\\
0 \quad&\text{else}\,,
\end{cases}
\end{equation}
we can perform the integration over the angles in spherical polar coordinates to give
\begin{equation}
U(r) = 2\pi \hbar c\alpha\int_{r'} \funop{n_n(r')} r'^2 \funop{\mathcal{I}(r, r')} \dd r'\,,
\end{equation}
where
\begin{equation}
\!\!\!\mathcal{I} \equiv \frac{\lambda}{r r'} \times
\begin{cases}
e^{(r - r')/\lambda} - e^{-(r+r')/\lambda}  &\text{if $r \leq r'$,}\\
e^{-(r - r')/\lambda} - e^{-(r+r')/\lambda}  &\text{if $r> r'$.}
\end{cases}
\end{equation}
The above integration can be performed analytically to give the potential energy $U(r)$, from which we can compute the force via $\bv{F} = -\nabla U(r)$. This yields
\begin{widetext}
\begin{align}
 U(r) &= \hbar c\frac{3\alpha}{(R_\mrm{eff} / \lambda)^3} \times
\begin{cases}
\frac{e^{-r/\lambda}}{r}\left[\frac{R_{\mathrm{eff}}}{\lambda} \cosh\left(\frac{R_\mrm{eff}}{\lambda}\right) - \sinh\left(\frac{R_\mrm{eff}}{\lambda}\right)\right]&\text{if $r \geq R_\mrm{eff}$}\,,\\
\frac{1}{\lambda} + \frac{1 + R_\mrm{eff}/\lambda}{r}\sinh\left(\frac{r}{\lambda}\right)\left[\sinh\left(\frac{R_\mrm{eff}}{\lambda}\right) - \cosh\left(\frac{R_\mrm{eff}}{\lambda}\right)\right]&\text{else}\,.
\end{cases}\\[0.5cm]
\bv{F}(\bv{r}) &= \hbar c\frac{\alpha}{r^2} \hat{\bv{r}} \times
\begin{cases}
e^{-r/\lambda}\left( 1 + \frac{r}{\lambda} \right) \mathcal{G}_1\left( \frac{R_{\mathrm{eff}}}{\lambda} \right)&\text{if $r \geq R_\mrm{eff}$}\,,\\
\left[ \frac{r}{\lambda} \cosh\left( \frac{r}{\lambda} \right) -\sinh\left( \frac{r}{\lambda} \right)\right] \mathcal{G}_2\left( \frac{R_{\mathrm{eff}}}{\lambda} \right)&\text{else}\,.
\end{cases}
\label{eq:yukawa_force}
\end{align}
\end{widetext}
Here, each function $\mathcal{G}$ is a geometric factor accounting for the finite size of the magnet:
\begin{equation}
\begin{split}
 \mathcal{G}_1(x) &\equiv \frac{3}{x^3}(x \cosh x - \sinh x)\,,\\
 \mathcal{G}_2(x) &\equiv \frac{3(1+x)}{x^{3}}(\cosh x - \sinh x)\,.
 \end{split}
\end{equation}
Ultimately, we consider trajectories external to the magnetic particle, such that only $\mcal{G}_1$ is relevant for our analysis.

In our DM search, we consider the momentum transferred by a passing DM particle to one of the translational modes of our magnetically levitated sensor. We approximate the DM trajectory to be a straight line $\bv{r}(t) = \bv{b} + \bv{v} t$, where $\bv{b}$ is the impact parameter vector, $\bv{v}$ is the DM velocity, and $t$ is the time measured from the point of closest approach. The impulse delivered to the sensor is then
\begin{equation}
 \bv{q}_\perp \equiv\int_{-\infty}^\infty \bv{F}\textbf{(}\bv{r}(t)\textbf{)} \dd t\,,
    \label{eq:mom_transfer}
\end{equation}
which is purely transverse and directed along $\hat{\bv{b}}$ since the longitudinal component cancels by symmetry. The straight-line treatment is valid as long as we are in the small-angle scattering regime, where the scattering angle satisfies $\Theta \ll 1$. 
In the center-of-mass frame of the DM--magnet system, the scattering angle and the magnitude of the transferred momentum are related by
\begin{equation}
 q_\perp(\Theta) = 2 \mu v\sin\left(\frac{\Theta}{2}\right) \simeq m_\mrm{DM} v \Theta\,,
 \label{eq:small_angle}
\end{equation}
where we have taken the reduced mass to be $\mu \simeq m_\mrm{DM}$ since $m_\mrm{mag} \gg m_\mrm{DM}$ and assumed $\Theta \ll 1$. We comment on how we satisfy this criterion at the end of this section.

For DM trajectories external to the magnetic particle, \cref{eq:mom_transfer} can then be evaluated analytically for the Yukawa force given in \cref{eq:yukawa_force}. The magnitude of the transferred momentum is then
\begin{equation}
    q_\perp(b) = \hbar c\frac{2  \alpha}{\lambda v}  \funop{K_1\left(\frac{b}{\lambda}\right)}\funop{\mcal{G}_1\left(\frac{R_\mrm{eff}}{\lambda}\right)}\,,
    \label{eq:mom_transfer_perp}
\end{equation}
where $K_1$ is the modified Bessel function of the second kind and of order one. While $\mcal{G}_1$ captures the coherence loss for a finite force range and sensor size, $K_1$ accounts for the exponential Yukawa suppression factor. In the long-range limit, where $\lambda \gg b \geq R_\mrm{eff}$, $K_1(x) \rightarrow 1/x$ and $\mcal{G}_1 \rightarrow 1$, such that we recover the Coulomb result $q_\perp \rightarrow 2 \hbar c \alpha / (bv)$.

In our analysis, the measured quantity $q$ is the impulse along a single sensitivity axis $\hat{\boldsymbol{\zeta}}$, such that $q(b) = q_\perp(b) |\hat{\bv{b}} \cdot \hat{\boldsymbol{\zeta}}|$. Since DM particles can pass by the detector with any relative orientation to the sensitivity axis, the same $q$ can be induced by a range of impact parameters. For a DM trajectory perpendicular to this axis, for which $|\hat{\bv{b}} \cdot \hat{\boldsymbol{\zeta}}| = 1$, the impact parameter for a given $q$ is maximized. All other trajectories carry a smaller projection factor and must approach the sensor more closely to deliver the same impulse. The differential cross section therefore marginalizes over all impact parameters producing the same projected transfer.

\begin{figure}[t]
 \centering
 \includegraphics{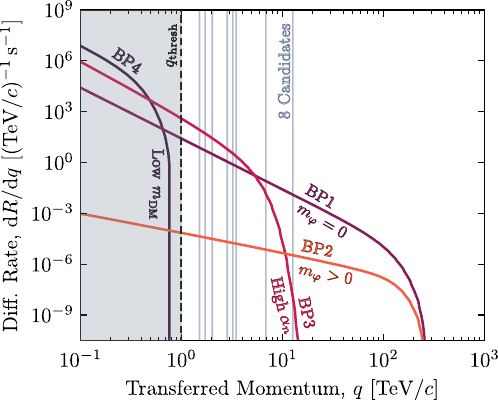}
 \caption{The differential rate $\dvil{R}{q}$ with momentum transfer $q$. Four benchmark points (BPs) are shown with default parameters $\alpha_n = 10^{-5}$, $m_\varphi = 0\,\mrm{eV}/c^2$, and $m_\mrm{DM} = 10^8\,\mrm{GeV}/c^2$, with one parameter changing at a time. BP1 features a massless mediator; BP2 a massive mediator with $m_\varphi = 1\,\mrm{meV}/c^2$ ($\lambda \approx 200\,\mrm{\mu m}$); BP3 a high coupling $\alpha_n = 8.2 \times 10^{-3}$, causing significant atmospheric attenuation; and BP4 a low DM mass $m_\mrm{DM} = 3\times 10^{5}\uni{GeV}/c^2$, resulting in momentum transfers that are always below our threshold. The dashed line shows our threshold momentum $q_\mrm{thresh} = 1\,\mrm{TeV}/c$, below which dark matter events are invisible to our detector.  The blue vertical lines mark the observed candidate impulses of mode~1. No detection efficiency is folded in.}
 \label{fig:sm-spectra}
\end{figure}

To perform this marginalization, we need the distribution of the projection factor $|\hat{\bv{b}} \cdot \hat{\boldsymbol{\zeta}}|$. For an isotropic DM flux, $\hat{\bv{b}}$ is uniformly distributed on the unit sphere, and its projection onto a fixed axis is also uniform. Then $|\hat{\bv{b}} \cdot \hat{\boldsymbol{\zeta}}| \sim U(0, 1)$, such that $q$ is also uniformly distributed as $q \sim U(0, q_\perp)$. This treatment ignores the anisotropy of the DM flux induced by the motion of the Sun around the DM halo, which we do not expect to have a significant impact on our results. In making this approximation, we mainly discard directional information, which we should also be sensitive to and which is studied more closely in the context of daily modulation in the optical levitation DM search of Ref.~\cite{Tseng:2025rlo}.

The differential cross section then follows from the geometric measure $\dd \sigma = 2\pi b \dd b$ weighted by the distribution of the projection factor. This classical scattering approach is valid as long as the de Broglie wavelength of the system $\lambda_\mrm{dB} \equiv \hbar / (\mu v)$, with $\mu \simeq m_\mrm{DM}$, is much smaller than the scale of the Yukawa potential $\lambda$. For all of our parameter space, we find that $\lambda_\mrm{dB}/\lambda \ll 1$. Since $q_\perp(b)$ decreases monotonically with $b$, the largest impact parameter that leads to a given projected impulse $q$ is $b_\mrm{max}(q)$, found by inverting \cref{eq:mom_transfer_perp}. The differential cross section is then given by
\begin{equation}
    \dv{\sigma}{q}(q) = 2 \pi \int_{R_\mrm{eff}}^{b_\mrm{max}(q)} \frac{b}{q_\perp(b)} \dd b\,,
\end{equation}
where the lower limit is set to $R_\mrm{eff}$ to exclude trajectories crossing our sensor. Since the highest momentum transfers occur at grazing incidence and interior trajectories only contribute an order-unity correction, their omission is conservative. For a massless mediator ($\lambda \rightarrow \infty$), we find
\begin{equation}
    \dv{\sigma}{q} = (\hbar c)^2\frac{1}{3} \frac{8\pi  \alpha^2}{v^2 q^3}\,,
\end{equation}
which is the Rutherford result with an additional projection factor of $\langle |\hat{\bv{b}} \cdot \hat{\boldsymbol{\zeta}}|^2\rangle = 1/3$, accounting for the projection onto one of our three translational axes.

The small-angle condition given in \cref{eq:small_angle} is maintained when we compute the differential rate expression in \cref{eq:diff_rate}
by only integrating over velocities greater than $v_\mrm{min}(q) \equiv q/m_\mrm{DM}$. This bounds
the scattering angle of the trajectories dominating the cross section. Violations of this condition are confined to trajectories passing close to the sensor, when $q_\perp(b) > q$, and
the contribution from these trajectories to the cross section is suppressed as
$(q/m_\mrm{DM}v)^3$. Therefore, our small-angle approximation degrades only in the low-mass corner where $q \rightarrow m_\mrm{DM}(v_\mrm{esc} + v_\odot)$; however, we expect uncertainties in quantities like the escape and solar velocities to be more important in this limit. For a general treatment valid for any scattering angle and both internal and external sensor trajectories, one should solve the classic orbital integral problem~\cite{Goldstein:2002mechanics}.

Combining the detector velocity distribution with the cross section result, we may compute the differential rate given in \cref{eq:diff_rate}. \Cref{fig:sm-spectra} shows the differential rate for four benchmark points (BPs), varying in the coupling, mediator mass and dark matter mass, with the eight candidates surviving our cuts overlaid. BP1 serves as the standard of comparison, showing the rate for a massless mediator and a DM mass of $m_\mrm{DM} = 10^{8}\,\mrm{GeV} / c^2$. For a massive mediator of $m_\varphi = 1\,\mrm{meV}/c^2$ ($\lambda \approx 200\,\mrm{\mu m}$), the cross section is suppressed at low momenta, decreasing the rate for BP2. For sufficiently high couplings, atmospheric attenuation effects suppress the rate and lower the kinematic floor to lower momenta due to the smaller accessible velocities, as shown for BP3 for $\alpha_n = 8.2 \times 10^{-3}$. Finally, we show that for DM masses $m_\mrm{DM} \lesssim q_\mrm{thresh} / (v_\mrm{esc} + v_\odot) \sim 4 \times 10^5\,\mrm{GeV}/c^2$, shown for BP4 for $m_\mrm{DM} = 3 \times 10^{5}\,\mrm{GeV}/c^2$, no events can be produced with momenta over our threshold of $q_\mrm{thresh} = 1\,\mrm{TeV}/c$.

\section{Impulse Search}
\label{sec:impulse-search}

We read out a single mechanical mode, and the fitted quantity is an amplitude in units of the demodulated stream. We calibrate the amplitude fitted to the most significant sample of each above-threshold excursion, and three constants convert that amplitude to a momentum. The demodulated voltage is first mapped to a displacement of the levitated magnet along the mode coordinate using the response of the calibration loop of Ref.~\cite{Uitenbroek:2026ybe}, which fixes the volts-per-meter scale of the readout at the mode frequency. A step $\delta x$ in the displacement of a freely evolving resonator corresponds to a velocity change $\delta v = 2\pi f_0\,\delta x$, such that a fitted kick amplitude becomes a velocity change on multiplication by $2\pi f_0$. Multiplying by the moving mass of the mode gives the momentum delivered to the sensor, which is the quantity we report for every candidate and the one in which the search threshold and the efficiency curve are expressed. No other factor enters; in particular, the reconstructed momentum is the component projected onto the mode coordinate, and we apply no angular acceptance correction for the transverse components the mode does not sense.

One systematic uncertainty matters for this chain. Ref.~\cite{Uitenbroek:2026ybe} quotes a relative uncertainty of $8.5\%$ on the displacement calibration, dominated by the uncertainty on the mass of the sensor. It enters the momentum scale linearly, so a shift of the calibration simultaneously slides the candidate spectrum, the momentum threshold, and the efficiency curve rather than distorting any of them relative to each other. We do not propagate this uncertainty into our exclusion, and we instead report the limits using the nominal calibration. The effect of an alternative scale can be read off by rescaling the momentum axis of \cref{fig:data-spectrum} and the predicted spectra by the same factor. 

To identify transient impulse candidates, we follow a log-likelihood approach. The data of the search are the first differences $d_k \equiv z_{k+1} - z_k$ of the demodulated complex amplitude $z$ of one mode, decimated and not averaged, onto the uniform $2.0\uni{s}$ grid. A contiguous segment carries $M$ such differences (with $M = 271$ when full), and the index $i$ labels the difference within its segment. The response of the lock-in filter to a step spans a time comparable to the grid spacing, so a kick generically splits between two adjacent differences. For a candidate kick shared between differences $i$ and $i{+}1$  , we model the $M$ complex difference samples as independent, circularly-symmetric Gaussian
variates, $d_k \sim \mcal{CN}(s_k, \sigma^2)$ with the variance
$\sigma^2\equiv\langle |d_k - s_k|^2 \rangle$. The log-likelihood for a kick at position $i$ is then
\begin{equation}
\begin{split}
\ln\mcal{L}_i(\delta v, w, \sigma^2) &= -M\ln \pi\sigma^2\\
 - \frac{1}{\sigma^2}\sum_{k=1}^{M}
   \big|&d_k - \delta v\left[w\,\delta_{ki} + (1-w)\,\delta_{k(i+1)}\right]\big|^2,
\end{split}
 \label{eq:transient-lnL}
\end{equation}
where $\delta v \in \mathbb{C}$ is the kick amplitude and $w \in [0,1]$ is the fraction
shared into the first difference. 

We use this log-likelihood to define the transient test statistic $\mcal{T}$ as the ratio of the log-likelihood under the impulse-free hypothesis to its value at the best fit:
\begin{equation}
\begin{split}
 \mcal{T}_i
  &\equiv -2\ln\left[\frac{\mcal{L}_i\!\left(\delta v = 0,\, \hat{\hat{\sigma}}^2\right)}
               {\mcal{L}_i\!\left(\hat{\delta v},\, \hat{w},\, \hat{\sigma}^2\right)}\right]\\
  &= -2M\ln\!\left(1 - \frac{R_\mrm{max}(i)}{S}\right),
\end{split}
 \label{eq:transient-closed}
\end{equation}
where in the second line we have analytically profiled all parameters under each hypothesis. Moreover, we have defined $S \equiv \sum_k |d_k|^2$ and
\begin{equation}
 R_\mrm{max}(i)
 \equiv \max_{w \in [0,1]} \left[
   \frac{\left|w\,d_i + (1-w)\,d_{i+1}\right|^2}{w^2 + (1-w)^2}\right]\,,
 \label{eq:transient-rmax}
\end{equation}
where $R_\mrm{max}$ is the recovered signal power maximized over the sharing fraction with the
amplitude and phase of the kick fitted at each $w$. We do not correct for the ringdown between samples in computing this recovered signal power as the
ringdown time $1/\gamma$ of the mode exceeds the $2.0\uni{s}$ sample spacing by approximately three orders of magnitude. Because the kick amplitude and phase are profiled, $\mcal{T}_i^{1/2}$ coincides with the matched-filter signal-to-noise ratio of the recovered kick amplitude in the small-signal limit $\mcal{T}_i \ll 2M$, and we quote it as such. The $\mcal{T} > 100$ selection used in our analysis then corresponds to a signal-to-noise ratio of $10$.

The null distribution of $\mcal{T}$ is known in closed form. Under the noise-only hypothesis, the differences $d_k$ are independent complex Gaussians, and the bracketed ratio of \cref{eq:transient-rmax}, $R_i(w)$, is a Rayleigh quotient of the $2\times2$ real Gram matrix $A$ built from $(d_i, d_{i+1})$, with $A_{11} = |d_i|^2$, $A_{22} = |d_{i+1}|^2$, and $A_{12} = \mrm{Re}\,(d_i d_{i+1}^{*})$. Maximizing over $w \in [0,1]$ therefore requires no grid scan: $R_\mrm{max} = \lambda_\mrm{max}(A)$ when $A_{12} \geq 0$, and $R_\mrm{max} = \max(A_{11}, A_{22})$ otherwise, the sign of $A_{12}$ being independent of all else and equally likely either way. The vector defined by components $|d_k|^2/S$ is Dirichlet-distributed and independent of $S$, so $\mcal{T}$ is scale free (the noise level cancels and the null needs no calibration to the data), and the probability $\Pr[\mcal{T} > t]$ reduces to a single two-dimensional quadrature over the relative amplitude and phase of the pair, with the boundary branch summing to a finite-$M$ correction of the $\chi^2_2$ tail.

We evaluate the selection threshold analytically, checking it against $4.0\times10^{9}$ noise-only per-index draws, simulated as whole segments and pooled exactly as the data are pooled so that both correlations of the analysis are preserved. Our analytics and numerics agree to within $1.4\sigma$ at every threshold the Monte Carlo resolves, down to $\Pr[\mcal{T} > t] = 6\times10^{-9}$ at $t = 41.2$. At the $\mcal{T} > 100$ selection, the exact per-index false-positive probability is $1.8\times10^{-21}$, giving $4.4\times10^{-15}$ expected Gaussian-noise accidentals over every per-index trial of all three recorded modes. The measured distributions follow this null through the bulk and depart from it far below the selection threshold. Non-Gaussian transients therefore limit the search, and we set the threshold where the exact Gaussian tail is negligible.

The detection efficiency $\varepsilon(q)$ of the $\mcal{T} > 100$ selection is evaluated analytically segment by segment, from the non-central $\chi^2$ distribution of the matched-filter power at the measured noise level of that segment, and then averaged over the selected segments. The three degrees of freedom are those of the kick amplitude, its phase, and the sharing fraction fitted at each pair; moving between two and three degrees of freedom shifts the momentum scale at the $50\%$ point by $0.46\%$. The efficiency is marginalized over the arrival phase of the impulse within the sample pair.

\section{Limit Setting}
\label{sec:limit-setting}

We set limits with the optimum-interval method of Ref.~\cite{Yellin:2002xd}, also used by the levitated-nanosphere search of Ref.~\cite{Tseng:2025rlo}. The underlying principle is that a region of the observable that is anomalously empty relative to a proposed signal spectrum bounds that signal from above regardless of the background.

A signal hypothesis $\mcal{H} \equiv (\alpha_n, \lambda, m_\mrm{DM})$ enters through the detected impulse spectrum. The expected number of detected impulses under the hypothesis $\mcal{H}$ then follows from the differential rate given in \cref{eq:diff_rate}:
\begin{equation}
 N_\mrm{exp}(\mcal{H}) = T_\mrm{obs}\int_{q_\mrm{thresh}}^{q_\mrm{max}} \varepsilon(q)\dv{R}{q}(q;\mcal{H})\dd q\,.
 \label{eq:mu_expected}
\end{equation}
Here, $T_\mrm{obs} \approx 9.15\,\mrm{days}$ is the experimental livetime, which is computed from the mode-1 lock-in output and taken to be common to all three modes; both the livetime and the efficiency are night-selection quantities, evaluated on the same vetoed, night-time segments the search runs on. The upper limit $q_\mrm{max}$ is the closest-approach endpoint described in \cref{subsec:cross_section}, above which the hypothesis predicts no rate. The lower endpoint $q_\mrm{thresh} = 1\uni{TeV}/c$ both regulates the massless $\dvil{\sigma}{q} \propto q^{-3}$ integral and defines the search: it lies below every candidate and sets the kinematic wall at $q_\mrm{thresh}/(v_\mrm{esc}+v_\odot) \approx  4\times10^{5}\uni{GeV}/c^2$. The efficiency at this point is $17\%$, reaching $50\%$ at $1.21\uni{TeV}/c$. The efficiency $\varepsilon(q)$ is folded into the differential rate, and the curve for mode 1 is shown in \cref{fig:data-spectrum}. The curves for modes~2 and~3, together with the transient up-crossings from which each mode's candidate list is drawn, are included in the data release~\cite{luhdm_datarelease}.

Candidates beyond $q_\mrm{max}$ cannot be a signal for that hypothesis and are dropped. From the survivors, mapped through the cumulative detected spectrum, we take the interval of the observable that is most anomalously empty relative to the signal $\mcal{H}$ predicts and reduce it to a single statistic, $C_\mrm{max}^\mrm{obs}(\mcal{H})$; this construction follows Ref.~\cite{Yellin:2002xd} and is set out step-by-step in the data release~\cite{luhdm_datarelease}. We calibrate the distribution of $C_\mrm{max}$ by Monte Carlo. Exclusion uses the quantile of the observed statistic in the simulated distribution,
\begin{equation}
 p(\mcal{H}) = \Pr\!\left[\,C_\mrm{max} < C_\mrm{max}^\mrm{obs}(\mcal{H}) \mid N_\mrm{exp}\right] \,,
 \label{eq:extremeness}
\end{equation}
where the probability is conditioned on background-free pseudo-experiments: $\mrm{Poisson}(N_\mrm{exp})$ impulses are drawn from the detected spectrum of $\mcal{H}$, and each is put through the same construction. We call $p$ the extremeness of $\mcal{H}$ and exclude hypotheses with $p \geq 0.95$; a large $p$ means the data are anomalously empty under $\mcal{H}$. Because a real background can only add candidates, shrink the intervals, and lower $C_\mrm{max}$, the rule ``exclude where $p \geq C$'' has coverage of at least $C$; we therefore quote our exclusions as at least $95\%$ CL. The quantity $p$ is a frequentist tail probability under the signal hypothesis and should not be taken as  a discovery $p$-value.

The scan evaluates \cref{eq:extremeness} independently at every point of the $(\alpha_n,\lambda, m_\mrm{DM})$ grid. At fixed mass and force range, the excluded set in coupling is bounded above as well as below since the coupling controls both the signal rate and the atmospheric attenuation. We locate both edges by root-finding the $p = 0.95$ crossing in $\log_{10}\alpha_n$, bit-consistent with the released grid. Beyond these shortcuts, our implementation in Ref.~\cite{optimum_interval} deviates from Ref.~\cite{Yellin:2002xd} only numerically, and identically for the data and the calibration. Its validation tests and calibration tables are released with the analysis code, and every excluded region reported here can be regenerated from the data release~\cite{luhdm_datarelease}.

The excluded regions are capped at DM mass by a flux argument. The expected number of DM particles passing within a
distance $b_\mrm{cap}$ of the sensor is
\begin{equation}
 N(m_\mrm{DM}) = f_\mrm{DM}\,\frac{\rho_\mrm{DM}}{m_\mrm{DM}}\,\langle v \rangle
 T_\mrm{obs}\,\pi b_\mrm{cap}^2\,,
 \label{eq:mcut}
\end{equation}
which falls with mass due to the falling DM number density. Requiring at least
$3$ transits within $b_\mrm{cap} = 10\uni{cm}$, the scale of the
trap, gives $m_\mrm{cut} \approx 8.4\times10^{14}\uni{GeV}/c^2$ at
$f_\mrm{DM} = 1$, and ten times lower at $f_\mrm{DM} = 0.1$. We claim no
exclusion above it: there the expected number of close transits during the
exposure falls below a few, so the limit would rest on a handful of encounters
rather than on a rate.

\section{Constraints on Composite Dark Matter}
\label{sec:fifth-force}

The composite dark matter candidate we consider is a bound state of total mass $m_\mrm{DM}$ formed from more fundamental constituents of mass $\bar{m}_d < m_\mrm{DM}$. As a composite particle, this candidate admits a finite spatial extent that can be expressed as~\cite{Coskuner:2018are}
\begin{equation}
\begin{split}
    R_\mrm{DM} &= \frac{\hbar}{c}\left(\frac{\mcal{N}_d}{4/3 \pi n_\mrm{Sat}}\right)^{1/3} \simeq \frac{\hbar}{c}\left(\frac{9\pi}{4} \frac{m_\mrm{DM}}{\bar{m}_d^4}\right)^{1/3} \\
    &\simeq 4 \uni{\mu m} \,\left(\frac{m_\mrm{DM}}{10^6 \uni{GeV}/c^2}\right)^{1/3}\left(\frac{1 \uni{keV}/c^2}{\bar{m}_d}\right)^{4/3}\,.
\end{split}
\end{equation}
Here, $\mcal{N}_d \simeq m_\mrm{DM} / \bar{m}_d$ is the number of dark constituents, $n_\mrm{sat} \simeq \bar{m}_d^3 / (3\pi^2)$ is their saturation density, and we have fiducialized to the benchmark we consider in \cref{fig:limits}. This illustrates that the characteristic scale of this candidate is smaller than both the force range we consider, $R_\mrm{DM} \ll 20 \uni{\mu m}$, and the effective radius of the magnet $R_\mrm{DM} \ll R_\mrm{eff} = 260 \uni{\mu m}$. This justifies the point-like approximation for the DM particle and therefore the coherence over it for our macroscopic sensor.

On the other hand, this same DM particle suffers a complete loss of coherence when considering interactions with individual nuclei. This is the relevant regime for DM direct detection experiments, for which the relevant scale to compare the DM size to is, at best, the inverse of the minimum momentum transfer to which these experiments are sensitive, $q_\mrm{min}$~\cite{Coskuner:2018are}. This is set by the threshold energy of these detectors, with $q_\mrm{min} \simeq (2 m_n E_\mrm{thresh})^{1/2} \sim 1 \uni{MeV}/c$ for a detector with an energy threshold $E_\mrm{thresh} \sim 1 \uni{keV}$. Thus, $q_\mrm{min} R_\mrm{DM}/\hbar \sim 10^{7} \gg 1$, and the DM particle interacts highly incoherently with individual nuclei. This dramatically weakens the bound that direct detection can set on the DM--nucleon cross section.

Concretely, the DM--nucleon cross section is proportional to the DM form factor as $\sigma_{\chi n} \propto |F_\mrm{DM}|^2$. Assuming that our composite particle can be described as a uniform-density sphere, this form factor takes the form $F_\mrm{DM}(x) = 3 j_1(x) / x$, with $j_1$ the order-one spherical Bessel function of the first kind. Inserting the above estimates, we find $|F_\mrm{DM}(q_\mrm{min} R_\mrm{DM}/\hbar)|^2 \sim 10^{-29}$.  Therefore, while DM direct detection experiments place leading constraints on the spin-independent WIMP--nucleon cross section of the order of $\sigma_{\chi n} \lesssim 10^{-48}\,\mrm{cm^2}$ at $m_\mrm{DM} = 40\,\mrm{GeV}/c^2$ at $90\%$ CL~\cite{LZ:2024zvo}, with comparable results from Refs.~\cite{XENON:2025vwd,PandaX:2024qfu}, the form factor from the finite size of the DM particle significantly suppresses the signal due to the loss of coherence. This pushes the constraint to $\sigma_{\chi n} \lesssim 10^{-19} \uni{cm^2}$ at $40\uni{GeV}/c^2$, and to $\usim10^{-14}\uni{cm^2}$ once the flux is rescaled to our mass range, far above our bound at $\sigma_{\chi n} \lesssim 10^{-27}\uni{cm^2}$.

Instead, the leading constraint in our parameter space of interest arises from a fifth force experiment searching for deviations from the gravitational inverse-square law. Fifth force experiments typically parametrize the new, spin-independent interaction via a Yukawa-like potential with respect to gravity. The total potential energy is
\begin{equation}
 U_5(r) = -\frac{G_N m_1 m_2}{r}\left(1 + \alpha_5 e^{-r/\lambda}\right)\,,
\end{equation}
where $G_N$ is Newton's constant and $\alpha_5$ parametrizes the strength of the new interaction relative to gravity. Compared to our parametrization,  $\alpha = G_N m_1 m_2 \alpha_5 / (\hbar c)$. Then, considering a new force that couples proportionally to neutrons and setting $\mcal{N}_{n,i} \simeq m_i/(2 u)$ for each of the interacting bodies, the neutron coupling $g_n$ can be expressed as
\begin{equation}
 g_n(\lambda) \simeq \sqrt{\frac{16\pi u^2 G_N \alpha_5(\lambda)}{\hbar c}}\,.
 \label{eq:g_n_alpha}
\end{equation}
For the force range scale of  $\lambda = 20\,\mrm{\mu m}$ that we consider in \cref{fig:limits}, the most constraining fifth force experiments are  torsion balance experiments. The most sensitive of these is the Washington (E\"{o}t-Wash) experiment of Ref.~\cite{Lee:2020zjt}, constraining values of  $\alpha_5(20 \uni{\mu m}) > 20$ and thus couplings of $g_n \gtrsim 2 \times 10^{-18}$. 

To compute the DM--neutron cross section, we begin from the total DM--neutron coupling given in \cref{eq:yukawa}. We take the number of dark constituents to be $\mcal{N}_d \simeq m_\mrm{DM} / \bar{m}_d$ and saturate the dark coupling to near the perturbative limit of $g_d = 1$, as in Refs.~\cite{Monteiro:2020wcb,Tseng:2025rlo}. Unlike $g_n$, $g_d$ is only weakly constrained, and $g_d \gg g_n$ is in any case required for an observable signal. Evaluating the fifth-force bound for our benchmark composite model, we find $\alpha_n \gtrsim 1.9\times10^{-6}$ at the DM mass where our limit sits furthest below it,  $m_\mrm{DM} = 1\times10^{7}\uni{GeV}/c^2$. Finally, converting this quantity with the same cross-section recast we apply to our own result constrains $\sigma_{\chi n} \gtrsim 2.1\times10^{-26}\,\mrm{cm^2}$. This boundary is more model-dependent than our own limit:  it moves in proportion to $g_d/\bar{m}_d$, neither of which is measured, while our search constrains $\alpha_n$ directly.

\begin{figure}[t!]
    \centering
    \includegraphics{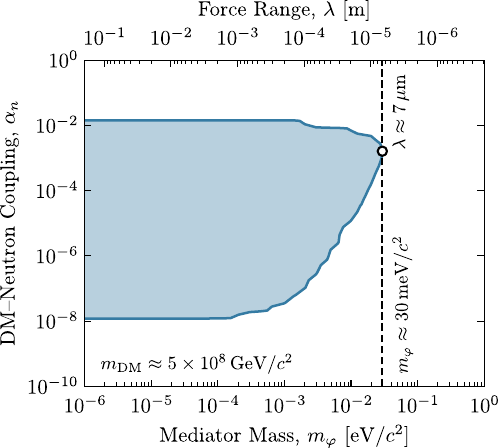}
    \caption{The 95\% CL limit on the DM--neutron coupling $\alpha_n$ with mediator mass $m_\varphi$ (bottom axis) and force range $\lambda$ (top axis). The dark matter mass is fixed at $m_\mrm{DM} \approx 5\times10^{8}\,\mrm{GeV}/c^2$, the value that optimizes this limit. The white circle marks the edge of the limit, occurring at the largest mass for which our experiment has sensitivity: $m_\varphi \approx 30\,\mrm{meV}/c^2$ ($\lambda \approx 7\,\mrm{\mu m}$).}
    \label{fig:lambda-limit}
\end{figure}

\section{Limit on Mediator Mass}

In \cref{fig:limits}, we show the limits on $\alpha_n$ with dark matter mass $m_\mrm{DM}$ for four mediator masses, stopping at the heaviest mass of $m_\varphi = 10\,\mrm{meV}/c^2$. However, we can also consider the limit at a fixed DM mass for varying mediator masses. By fixing the DM mass to that at which we have peak sensitivity, this limit tells us the heaviest mediator to which our experiment has sensitivity.

We show this limit in \cref{fig:lambda-limit}. The mass is held fixed at the optimal mass $m_\mrm{DM} = 5.0\times10^{8}\uni{GeV}/c^2$. The sensitivity closes off from below in $\alpha_n$ as the Yukawa
factor suppresses the signal for impact parameters $b > \lambda$. From above, the
atmospheric ceiling cuts off our sensitivity due to attenuation effects. The heaviest mediator still excluded is $m_\varphi \approx 30\,\mrm{meV}/c^2$ ($\lambda \approx 7\,\mrm{\mu m}$).

\section{\texorpdfstring{Analysis Code and Data Release}{Analysis Code and Data Release}}
\label{sec:data-release}

Every result in this work is produced using notebooks reading a single HDF5 file, which we release with the paper in Ref.~\cite{luhdm_datarelease} and which is also available on GitHub~\href{https://github.com/PolonaiseExperiment/luhdm.git}
{\faGithub}. On one grid over dark matter fraction $f_\mrm{DM}$, sensor mode, DM--neutron coupling $\alpha_n$, dark matter mass $m_\mrm{DM}$ and force range $\lambda$, it stores the optimum-interval extremeness $p$, the expected number of detected impulses $N_\mrm{exp}$, and the geometric transit count. A companion cube computed with atmospheric attenuation switched off, the measured per-mode efficiency curves, the candidate impulse lists and the livetime are stored in the same file. 

\vfill
\clearpage

\end{document}